\documentclass{svproc}
\usepackage{graphicx}%
\usepackage{multirow}%
\usepackage{amsmath,amssymb,amsfonts}%
\usepackage{mathrsfs}%
\usepackage[title]{appendix}%
\usepackage{xcolor}%
\usepackage{textcomp}%
\usepackage{manyfoot}%
\usepackage{booktabs}%
\usepackage{algorithm}%
\usepackage{algorithmicx}%
\usepackage{algpseudocode}%
\usepackage{listings}%
\usepackage{url}
\usepackage[table]{xcolor}
\usepackage{subcaption}
\usepackage{adjustbox}

\usepackage{soul}

\def\orcidID#1{\unskip$^{[#1]}$} 

\begin{document}
\mainmatter              
\title{No Community Detection Method to Rule Them All!}
\titlerunning{No Best Community Detection Algorithm}  
%
\author{Shrabani Ghosh\orcidID{0000-0002-6084-4964} \and Erik Saule\orcidID{0000-0003-1634-9234}
}
\authorrunning{Ghosh, Saule} 

\institute{University of North Carolina at Charlotte\\
\email{\{sghosh15, esaule\}charlotte.edu},
Charlotte, NC, USA}

\maketitle              

\begin{abstract}
Community detection is a core tool for analyzing large real-world graphs. It is often used to derive additional local features of vertices and edges that will be used to perform a downstream task, yet the impact of community detection on downstream tasks is poorly understood. Prior work largely evaluates community detection algorithms by their intrinsic objectives (e.g., modularity). Or they evaluate the impact of using community detection onto on the downstream task. But the impact of particular community detection algortihm support the downstream task.
We study the relationship between community structure and downstream performance across multiple algorithms and two tasks. Our analysis links community-level properties to task metrics (F1, precision, recall, AUC) and reveals that the choice of detection method materially affects outcomes. We explore thousands of community structures and show that while the properties of communities are the reason behind the impact on task performance, 
no single property explains performance in a direct way. Rather, results emerge from complex interactions among properties. As such, no standard community detection algorithm will derive the best downstream performance. 
We show that a method combining random community generation and simple machine learning techniques can derive better performance.
\end{abstract}

\section{Introduction}\label{intro}

Community detection is a crucial graph mining task that aims to reveal the structure of networks by identifying groups of nodes with strong interconnections. While community detection is sometimes used for its own direct benefit (for instance for visualization systems), they are often used to derive features for downstream tasks such as link prediction, node classification~\cite{ghosh2025-arXiv1}. 
Community detection proves useful for online privacy concerns~\cite{remy2018tracking} and identifying users based on their online behaviors. Waskiewicz~\cite{waskiewicz2012friend} illustrates that community detection can uncover those who endorse and propagate criminal ideas, potentially including individuals with terrorist affiliations. In bioinformatics~\cite{manipur2021community}, community detection allows for the confident inference of potential protein-protein interactions, the uncovering of gene regulatory networks, and the analysis of metabolic pathways; driving advancements in drug discovery, improve disease prediction, and enable personalized medicine. 

Several approaches to community detection have been proposed, and survey papers have provided a summary and comparative analysis of these methods, aiding in understanding the current state-of-the-art~\cite{rehman2012graph,su2022comprehensive,jin2021survey}.  However these studies usually focus on the impact of the algortihm on the metrics optimized by the community detection algorithms rather than on the impact of the algorithm on the application that they are part of.
We showed in a previous study that the choice of a community detection method for an application can significantly impact the downstream performance~\cite{ghosh2025-arXiv1}. 

In this paper, we study the mechanisms by which community detection impact the performance of downstream tasks. The intent is to enable application developers to pick better community structures for their application. As such we seek to answer the following research questions: How does structural properties impact downstream performance? Is there a good community detection method? How to select a community detection algorithm for an application? We restrict our study to non-overlapping community detection methods.

In Section~\ref{sec:cd}, we present background information about community detection, and we describe the two applications that we use in this study with the main results we derived in our previous study~\cite{ghosh2025-arXiv1}. Section~\ref{sec:propertyspace} shows how to generate many community structures using genetic algorithms. Analyzing the solutions generated by the genetic algorithm, we shows that properties traditionally used to evaluate the quality of community structures do not correlate strongly. Section~\ref{sec:downstreamspace} study the relation between the properties of community structures and the performance of the downstream analysis on the two application. The analysis highlights that the properties of community structure correlate weakly with the performance of the downstream analysis. These two results put together show that no traditional algorithm to optimize community structure should give absolutely good performance. Finally, Section~\ref{sec:perflearning} show that the properties of the community structure still can be used to predict the performance of the downstream analysis in ways that enable to identify which  community structures in a pool of structure are the most likely to give you good downstream performance without having to conduct a downstream performance analysis for each community structure.

\section{Community Detection}\label{sec:cd}

\subsection{Community detection methods}
A community detection (CD) method is an algorithm designed to identify groups of nodes, or communities, within a graph $G = (V,E)$, where $V$ is the set of nodes and $E$ is the set of edges. In general, these methods aim to partition or cluster the nodes such that the internal density of edges within each community $C_i \in V$ is higher than the density of edges between different communities. However, in practice the different methods are optimizing different objectives that are thought to derive good communities. Common objective functions include modularity, clustering coefficient, density, and conductance~\cite{Newman10book}.

We are restricting in this paper our study to non-overlapping community detection techniques, where each node belongs to a single cluster. The Louvain method~\cite{blondel2008fast} uses modularity maximization. Spectral methods~\cite{donath1973lower} uses eigenvectors to embed nodes in eigenspace and k-means clustering to derive communities. Label Propagation~\cite{raghavan2007near} uses neighboring nodes’ clustering information to propagate communities by greedily optimizing a quality metric. 
%
%
There are also other approaches based on embeddings such as Deepwalk~\cite{perozzi2014deepwalk}, AGE~\cite{cui2020adaptive} or GEMSEC~\cite{gemsec}.

Since different community detection methods focus on various optimization goals, they can produce widely different community structures. It is difficult to identify what makes a good community detection algorithm.
Even though some problem have ground truth communities, large datasets usually do not; and different algorithm seem to better recover communities for different datasets.

\subsection{Community Detection to Power Downstream Tasks}\label{sec:cddownstreamanalysis}

Applications that use community detection usually uses it as an intermediate step. They take the community structure generated by the Community Detection algorithm on some graph and pass it to further achieve the specific goals of the applications. Typically the communities are used to extract features used in a down stream machine learning model. We studied previously two applications of community detection~\cite{ghosh2025-arXiv1}: 1) Recommendation Systems that leverage community propensity information to generate personalized product recommendations
based on group interests; 2) Trust Prediction which utilize community structures
in social graphs to predict absent or potential link/trust relationships among users. 

\begin{figure}[t]
    \centering
    \includegraphics[width=0.9\linewidth]{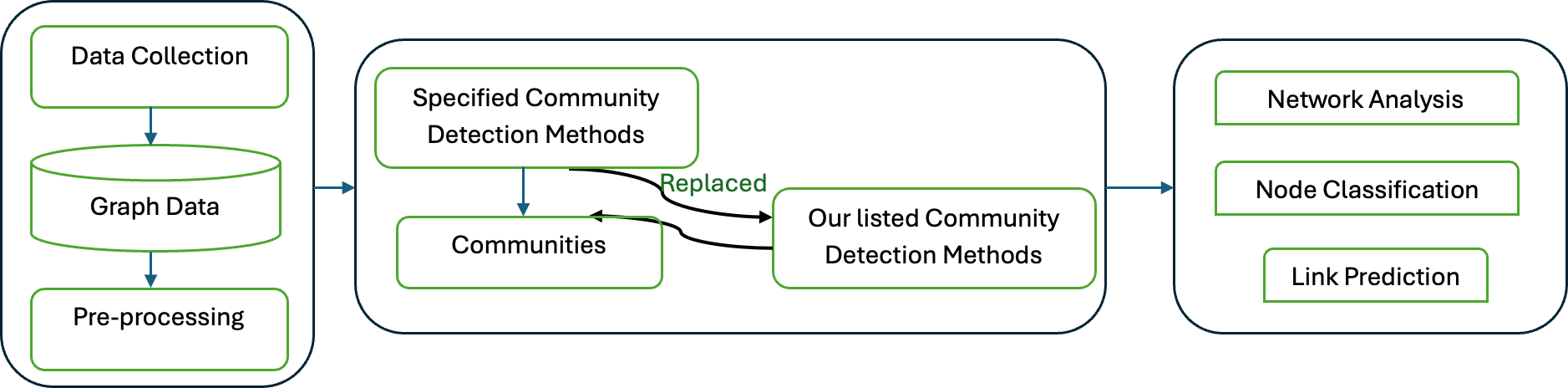}
    \caption{Framework to Study the Impact of Community Detection on an Application}
    \label{fig:framework}
\end{figure}

To study the impact of the community detection algorithm on existing applications, we followed a common framework summarized in Figure~\ref{fig:framework}. We replicate the methodology of the applications, but we replace the original community detection method that is used in the application with a standard set of selected methods and leave the rest of the methods untouched follow the rest of the process accordingly. We then evaluate how the application performs on metrics relevant to the application while changing the community detection algorithm used. We now describe the two application and the impact of community detection.

\subsection{Anomaly Detection by Keyvanpour et al.~\cite{keyvanpour2020ad}}\label{sec:app_anomaly}
Keyvanpour et al.~\cite{keyvanpour2020ad} studies the detection of anomalous node in a graph. They frame the problem as a node classification task on a graph assuming they have some ground truth. They utilize a social network graph of the reviews and its properties to detect anomalous nodes. 
Following the model described in the paper, they compute non-overlapping communities for the graph; then add auxilary communities for every edge connecting two different communities. They use six features of the nodes and communities to build their model: Number of communities the node belong to, the number of communities the node belongs to divided by its number of neighbors, 1 - cluster coefficient of the node, the degree of the node divided by the sum of its edge weights, the cliqueness of the node's main community, and the Starkness of the node's main community. Then they apply a thresholded linear regression to model the anomality of the node. The details of the method can be found in~\cite{ghosh2025-arXiv1}

\begin{table}[t]
\centering
\renewcommand{\arraystretch}{1.2}
\caption{YelpHotel anomaly detection (5-fold average)}
\label{tab:anomalydata_yelp}
\begin{adjustbox}{width=.8\textwidth}
\begin{tabular}{|l|l|c|c|c|c|c|c|}
\hline
Method & Class & Precision & Recall & F1 & Support & Accuracy & AUC \\ \hline\hline

\multirow{2}{.3\linewidth}{Single Community\\(Baseline)} & Normal  & 0.95 & 0.93 & 0.94   & 2035 & \multirow{2}{*}{90\%} & \multirow{2}{*}{0.62} \\ \cline{2-6}
                                  & Anomaly & 0.53 & 0.62 & 0.57 & 250  &                         &                      \\ \hline\hline

\multirow{2}{*}{Louvain}            & Normal  & 0.97 & 0.95 & 0.96 & 2035 & \multirow{2}{*}{93\%} & \multirow{2}{*}{0.71} \\ \cline{2-6}
                                    & Anomaly & 0.65 & 0.74 & 0.69 & 250  &                       &                       \\ \hline
\multirow{2}{*}{Spectral}           & Normal  & 0.97 & 0.96 & 0.96 & 2035 & \multirow{2}{*}{94\%} & \multirow{2}{*}{0.63} \\ \cline{2-6}
                                    & Anomaly & 0.69 & 0.75 & 0.72 & 250  &                       &                       \\ \hline
\multirow{2}{*}{Label Propagation}  & Normal  & 0.96 & 0.95 & 0.96 & 2035 & \multirow{2}{*}{92\%} & \multirow{2}{*}{0.64} \\ \cline{2-6}
                                    & Anomaly & 0.64 & 0.72 & 0.67 & 250  &                       &                       \\ \hline

\end{tabular}
\end{adjustbox}
\end{table}

We tested it in the YelpHotel dataset where a node is marked as suspicious when the associated reviewer has been identified for potentially submitting fake reviews. The results are in Table~\ref{tab:anomalydata_yelp}.
Grouping all nodes in a single community (aka, not conducting community detection) yields an accuracy of 90\%. The method classifies normal nodes reasonably well (precision 95\%, recall 93\%) but struggles to recognize anomalous nodes (Precision: 53\%, Recall 62\%). After applying community detection, all algorithms perform better than using a single community. The spectral method performs the best raising the precision and recall of the normal node by 2\% and of the anomalous nodes by 12\%.
Overall, this study shows community detection algorithm help in detecting anomalies on the YelpHotel dataset and can provide significant performance improvement.

\subsection{Trust Prediction by Beigi et al.~\cite{beigi2014leveraging}} \label{sec:app_trust}

Beigi et al.~\cite{beigi2014leveraging} predict trust between users in social networks using product ratings provided by leveraging users within the same community. When modeling trust as an edge between two users, trust prediction becomes and edge prediction task between user nodes. The model uses community information and the "center" of communities (identified using a centrality metric) as proxies for user behavior within that community. The details of the method can be found in~\cite{ghosh2025-arXiv1}.

\begin{table}[t]
\centering
\begin{minipage}{0.44\textwidth}
{\renewcommand\arraystretch{0.83} \footnotesize
\adjustbox{max width=\textwidth}{
\begin{tabular}{|c|c|c|c|c|} \hline
Centrality & Precision & Recall & F1 & AUC \\ \hline\hline
\multicolumn{5}{|c|}{Louvain} \\ \hline
Betweenness & 0.7997 & 0.8184 & 0.8089 & 0.5974 \\ \hline
MaxDegree   & 0.8021 & 0.8171 & 0.8095 & \textbf{0.6020} \\ \hline
MaxTrustor  & 0.7854 & 0.8515 & 0.8171 & 0.5721 \\ \hline
MaxTrustee  & 0.7942 & 0.8254 & 0.8095 & 0.5875 \\ \hline
Random      & 0.7862 & 0.8540 & 0.8187 & 0.5739 \\ \hline
\multicolumn{5}{|c|}{Spectral} \\ \hline
Betweenness & 0.7434 & 0.5344 & 0.6218 & 0.4867 \\ \hline
MaxDegree   & 0.7526 & 0.5756 & 0.6523 & 0.5001 \\ \hline
MaxTrustor  & 0.7465 & 0.5837 & 0.6552 & 0.4904  \\ \hline
MaxTrustee  & 0.7498 & 0.5632 & 0.6432 & 0.4958 \\ \hline
Random      & 0.7512 & 0.5867 & 0.6588 & 0.4979   \\ \hline
\multicolumn{5}{|c|}{Label Propagation} \\ \hline
Betweenness & 0.7192 & 0.6752 & 0.6965 & 0.4367  \\ \hline
MaxDegree   & 0.7192 & 0.6752 & 0.6965 & 0.4367 \\ \hline
MaxTrustor  & 0.7330 & 0.7115 & 0.7221 & 0.4618  \\ \hline
MaxTrustee  & 0.7295 & 0.7024 & 0.7157 & 0.4552 \\ \hline
Random      & 0.7349 & 0.7359 & 0.7354 & 0.4643   \\ \hline 
\end{tabular}}}
\end{minipage}
\begin{minipage}{0.44\textwidth}
{\renewcommand\arraystretch{0.85} \footnotesize
\adjustbox{max width=\textwidth}{
\begin{tabular}{|c|c|c|c|c|} \hline
Centrality & Precision & Recall & F1 & AUC \\ \hline\hline
\multicolumn{5}{|c|}{Louvain } \\ \hline
Betweenness & 0.6841 & 0.9370 & 0.7908 & 0.5362 \\ \hline
MaxDegree   &  0.6834 & 0.9432 & 0.7926 & 0.5351 \\ \hline
MaxTrustor  & \textbf{0.6968} & 0.9009 & 0.7858 & \textbf{0.5588} \\ \hline
MaxTrustee  & 0.6949 & 0.9038 & 0.7857 & 0.5556 \\ \hline
Random      & 0.6953 & 0.8992 & 0.7842 & 0.5560 \\ \hline
\multicolumn{5}{|c|}{Spectral} \\ \hline
Betweenness & 0.6789 & 0.6811 & 0.6800 & 0.5187 \\ \hline
MaxDegree   &  0.6792 & 0.6544 & 0.6666 & 0.5184 \\ \hline
MaxTrustor  & 0.6764 & 0.6904 & 0.6833 & 0.5153 \\ \hline
MaxTrustee  & 0.6775 & 0.6825 & 0.6800 & 0.5167 \\ \hline
Random      & 0.6741 & 0.6994 & 0.6865 & 0.5119 \\ \hline
\multicolumn{5}{|c|}{Label Propagation} \\ \hline
Betweenness & 0.6693 & 0.9852 & 0.7971 & 0.5064  \\ \hline
MaxDegree   & 0.6690 & \textbf{0.9863} & 0.7972 & 0.5057 \\ \hline
MaxTrustor  & 0.6695 & 0.9725 & 0.7931 & 0.5068  \\ \hline
MaxTrustee  & 0.6695 & 0.9852 & 0.7972 & 0.5068 \\ \hline
Random      & 0.6695 & 0.9756 & 0.7941 & 0.5067   \\ \hline 
\end{tabular}}}
\end{minipage}
\caption{Evaluation on Trust Prediction on Ciao (left) and Epinion (right) depending on the algorithm used for community detection and the centrality used to identify the "center" of the communities.}
\label{tab:trustnet_result}
\end{table}

Tables~\ref{tab:trustnet_result} show the results of predicting trust among users using rating information from the Ciao and Epinion datasets. The Louvain method performs better on the Ciao dataset, while Label Propagation performs better on the Epinion dataset. The centrality metric used to identify the "center" of the community does not have significant impacts. However the different community detection algorithm yield significant performance difference, with an increase in F1 score of 0.15 between the worst algorithm for the task (spectral) and the best algorithm for the task (Louvain).

\paragraph{Summary.} The analysis demonstrates that different community detection algorithms yield varying levels of performance on downstream tasks. There is no best algorithm for all tasks. However, it is hard to draw robust conclusions analyzing only few community detection algorithm as they represent a  small sample size.

\section{Study of Property Space of Communities}\label{sec:propertyspace}

To better understand which algorithm will perform best, we believe that we need to understand the community structure as seen through the lens of the properties of the community (their size, modularity, clustering coefficients, etc.).
To make the exploration possible we need a large sample size of good community structures, ideally thousands or more. Collecting thousands of community detection algorithms would be challenging and could suffer from a bias rooted in prefered type of techniques to build communities. We turn to Genetic Algorithm (GA) to generate a large and diverse set of community structure populations.

\subsection{Randomized Communities Generation} 

Genetic Algorithms systematically and randomly explores the solution space, enabling a more comprehensive examination of structural variations at scale. This approach allows us to study how various structural features co-occur, how populations differ when optimized under different objective functions, and to what extent solutions are similar or distinct within and across populations.
(There are other methods that could achieve our goals; we picked Genetic Algorithms because they are well studied and easy to use.) 

Genetic algorithms generate a large solution pool by starting with a randomly initialized set of solution. New solutions are created in the crossover stage which mixes two existing solution into a new solution. Solution can also be randomly perturbed (usually to get a better solution) in a mutation stage. Finally, a fitness functions is used to apply pressure to the solution pool.  We now describe the design choice we picked in our study.

\textbf{Initialization: }
We randomly initialized communities without considering any node connections. We picked a number of communities ranging from 20 to 160 communities. Each node was randomly assigned one community. 

\textbf{Fitness function:}
Genetic Algorithms need a driving function to optimize the solution and we considered different structural properties, both global properties such as the modularity~\cite{newman2006modularity} of the decomposition and local properties defined on particular communities and aggregated through averages such as and density~\cite{wasserman1994sna}, clustering coefficient~\cite{li2017clustering}, and conductance~\cite{leskovec2009community}.

\textbf{Crossover: } We randomly (uniformly) select four solutions from the population and sort them in descending order based on their fitness values. The top two solutions are chosen as parents. We then randomly select a community from the first parent and assign its nodes to the corresponding positions in the child solution. For the remaining nodes (those not yet assigned), we take their community information from the second parent.

\textbf{Mutation:} The mutation procedure starts by creating a copy of the original solution. Then, for each node in the graph, there is a small chance (probability of 0.1) that the node will be considered for mutation. If selected, the node's immediate neighbors in the graph are retrieved. If the node has any neighbors, their community assignments are examined, and the most frequently occurring community among them is identified. The node is then reassigned to that most common neighboring community. This process helps the solution evolve by making small, locally-informed adjustments. Finally, the modified solution is returned.

\textbf{Selection Process: }
After all solutions are generated, we select solutions in three ways.  We sort the solutions by fitness value in descending order. First, we select the top 20\% of the total solutions. Second, from the remaining solutions, we select 60\% of the solutions using a probabilistic Roulette Wheel Selection, where a community decomposition has a probability of being kept proportional to their fitness. This balances exploitation (using the best solutions) and exploration (trying out diverse or weaker solutions), helping the algorithm to search more effectively across the solution space. Third, from what remains, we randomly pick 100 solutions to maintain some diversity. If the total number of solution kept is higher than 100,000 we down sample the solution pool randomly uniformly.


\subsection{Analysis of Community Structure Properties}

\begin{figure}
    \centering
    \includegraphics[width=0.8\linewidth]{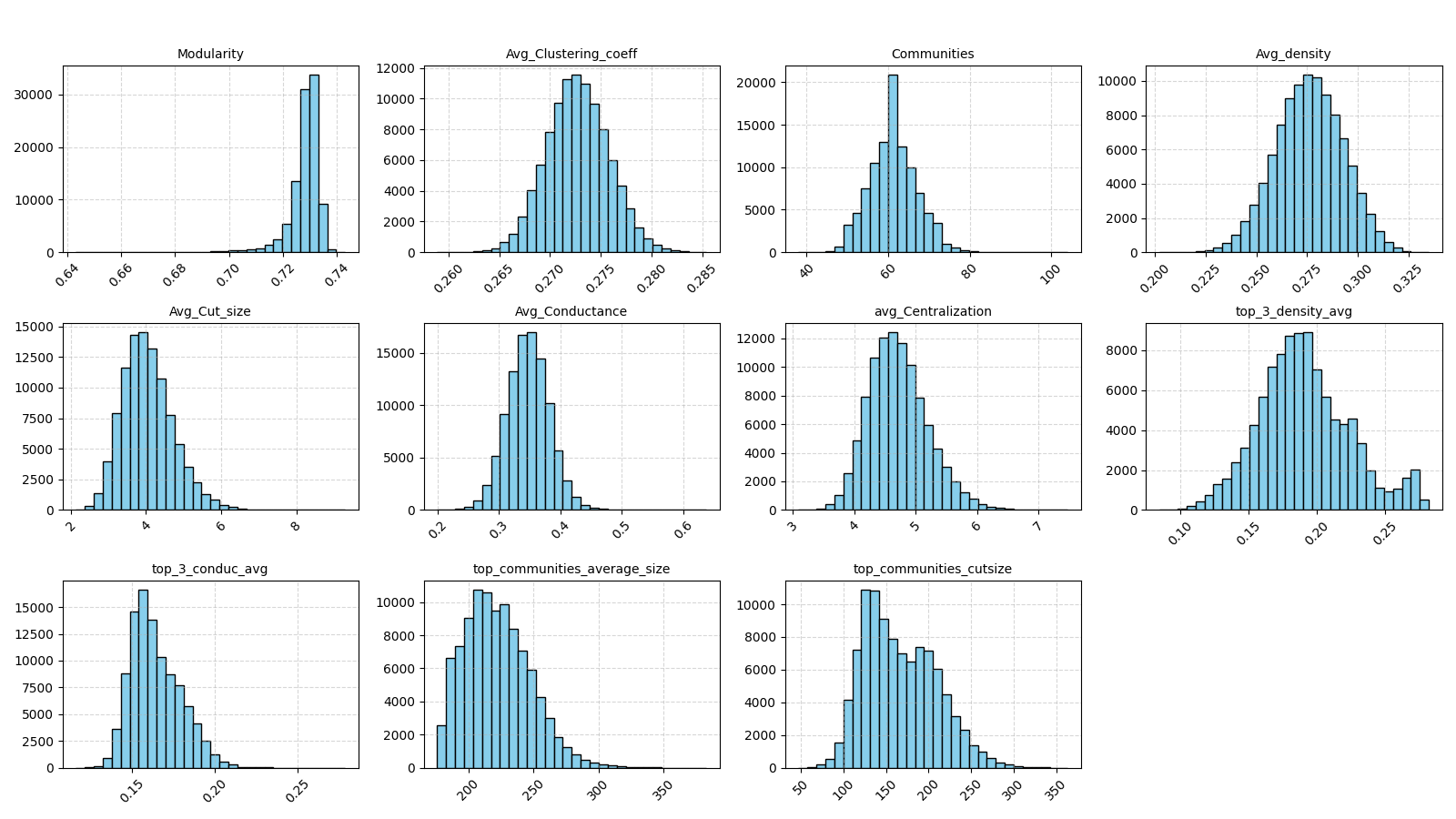}
    \caption{Modularity Optimized Solutions distribution of features: Yelp dataset}
    \label{fig:dist_yelp1}
\end{figure}

\begin{figure}
    \centering
    \includegraphics[width=0.8\linewidth]{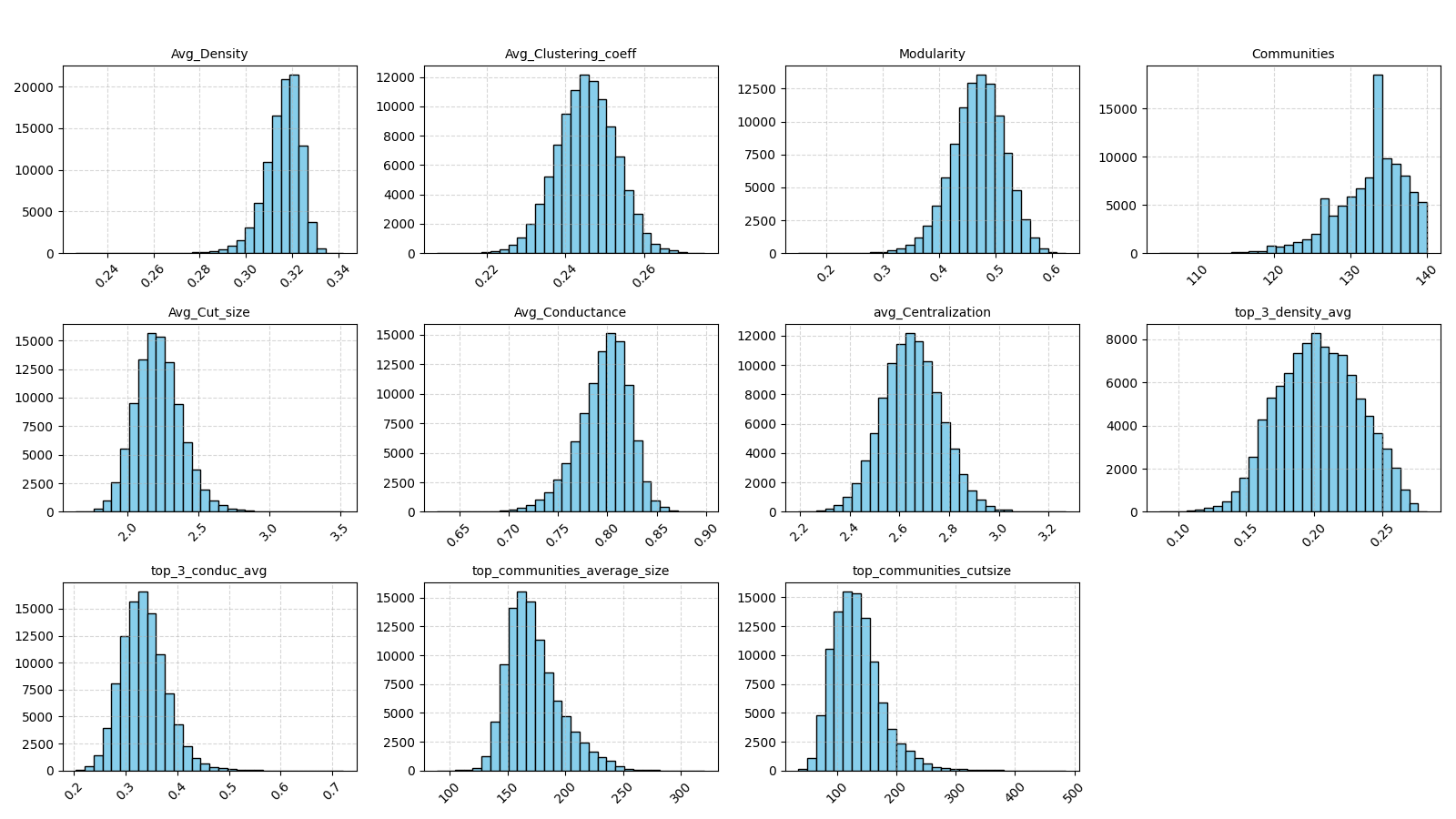}
    \caption{Density Optimized Solutions distribution of features: Yelp dataset}
    \label{fig:dist_yelp2}
\end{figure}

Different fitness function optimized solutions have different property distributions. We explored the distribution of properties to understand how they vary from each other. Figures~\ref{fig:dist_yelp1} and \ref{fig:dist_yelp2} show the distribution of properties when optimized by modularity and average density, respectively. In Figure~\ref{fig:dist_yelp1}, the median of modularity value is 0.73, number of communities is 60, and average clustering coefficient is 0.27. Conversely, Figure~\ref{fig:dist_yelp2} the median of modularity is 0.5, no of communities is 130 , and an average clustering coefficient of 0.24. 


For each solution we computed the following 11 properties: a) Modularity,
b) No of Communities,
c) Average Clustering  Coefficient,
d) Average Density,
e) Average Cut size,
f) Average Conductance,
g) Average Centralization,
h) Top 3 Communities Average Density,
i) Top 3 Communities Average Conductance,
j) Top Communities Average Size,
k) Top Communities Cutsize.  

\begin{figure}[t]
    \centering
    \begin{subfigure}[b]{0.48\linewidth}
        \centering
        \includegraphics[width=\linewidth]{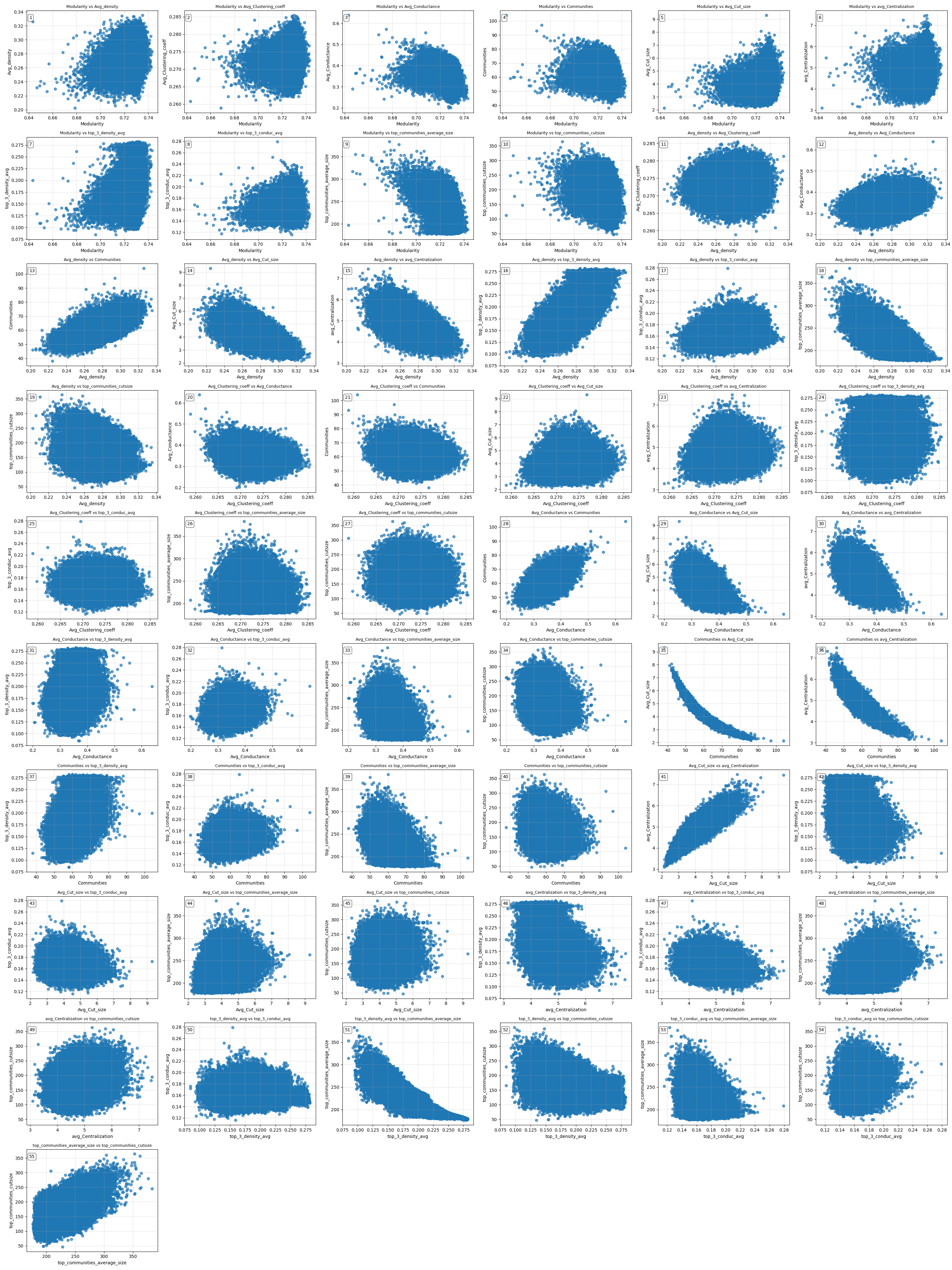}
        \caption{Modularity Optimized}
        \label{fig:yelp_mod8}
    \end{subfigure}
    \hfill
    \begin{subfigure}[b]{0.48\linewidth}
        \centering
        \includegraphics[width=\linewidth]{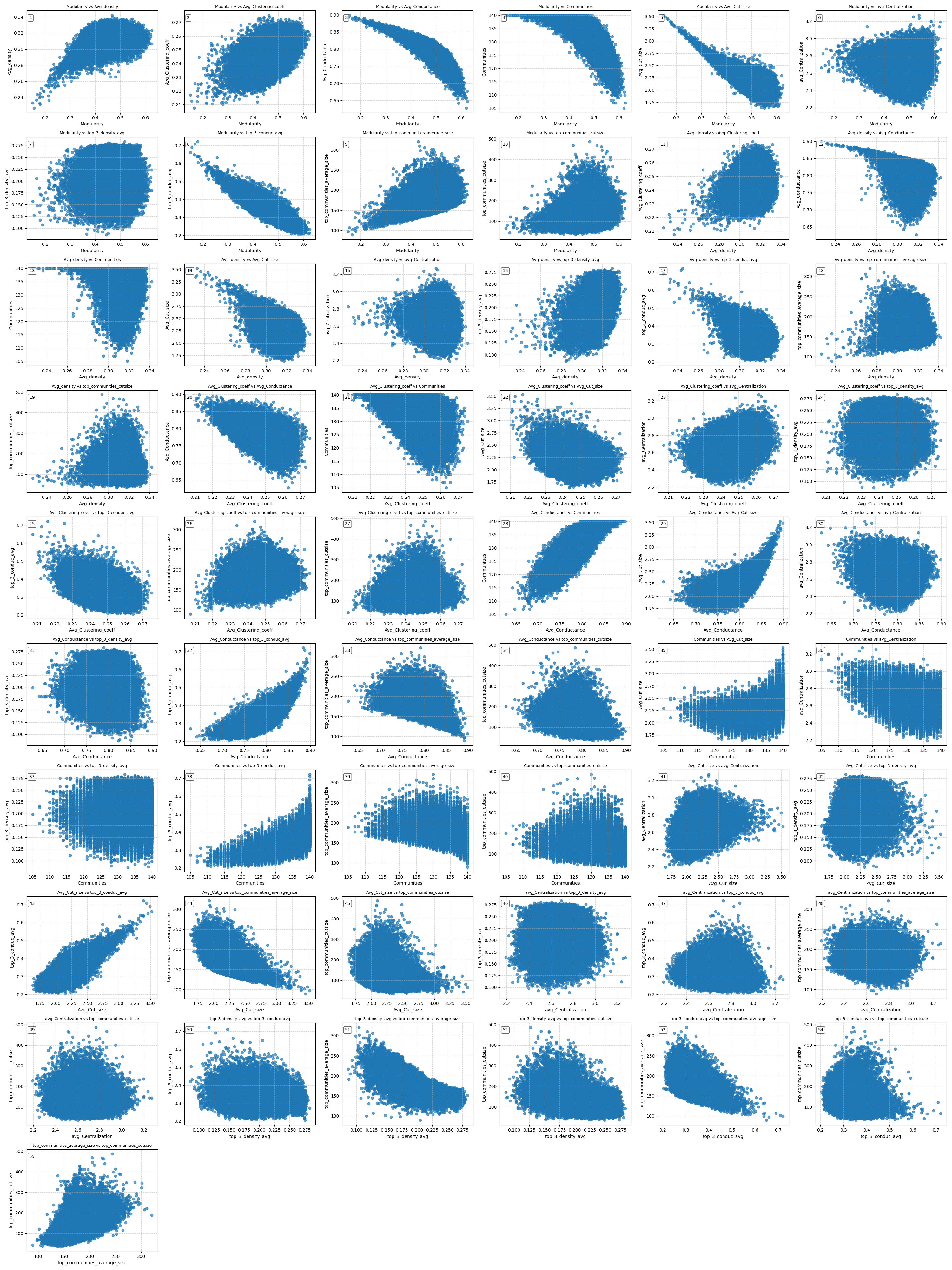}
        \caption{Density Optimized}
        \label{fig:yelp_den13}
    \end{subfigure}
    \caption{Comparison of feature correlations from modularity and density optimized solutions: YelpHotel Dataset}
    \label{fig:yelp_corr_comparison}
\label{fig:corr}
\end{figure}

\begin{figure}[t]
    \centering
    \begin{subfigure}[b]{0.48\linewidth}
        \centering
        \includegraphics[width=\linewidth]{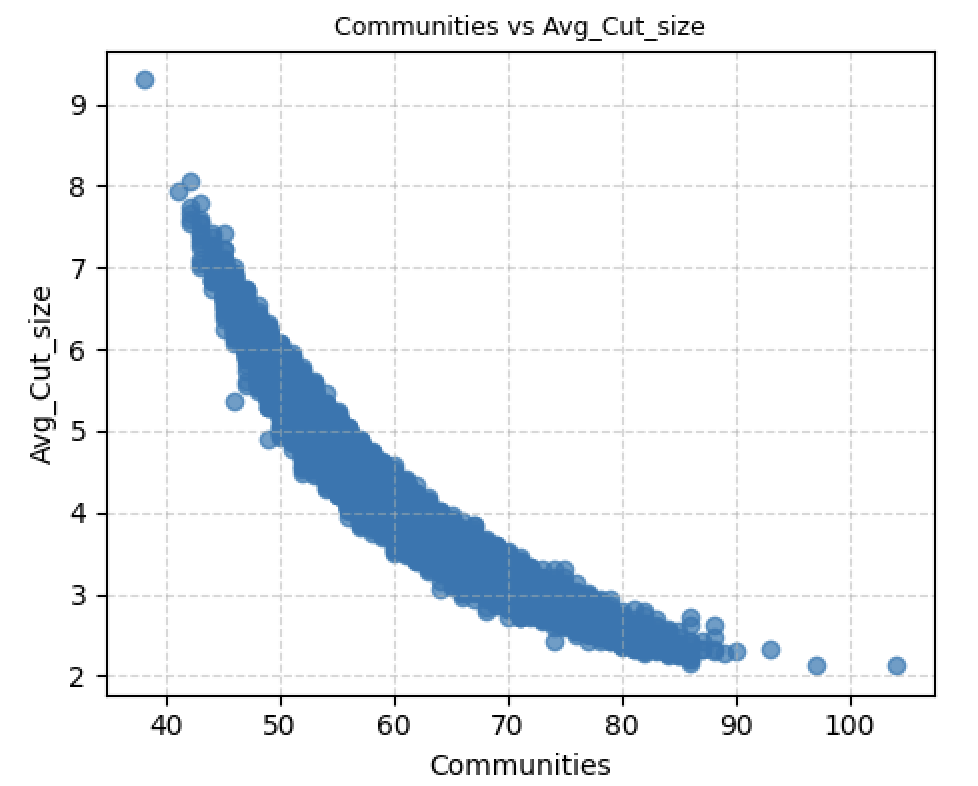}
        \caption{No of Communities vs Avg Cut size }
        \label{fig:subplot_modopt1}
    \end{subfigure}
    \hfill
    \begin{subfigure}[b]{0.48\linewidth}
        \centering
        \includegraphics[width=\linewidth]{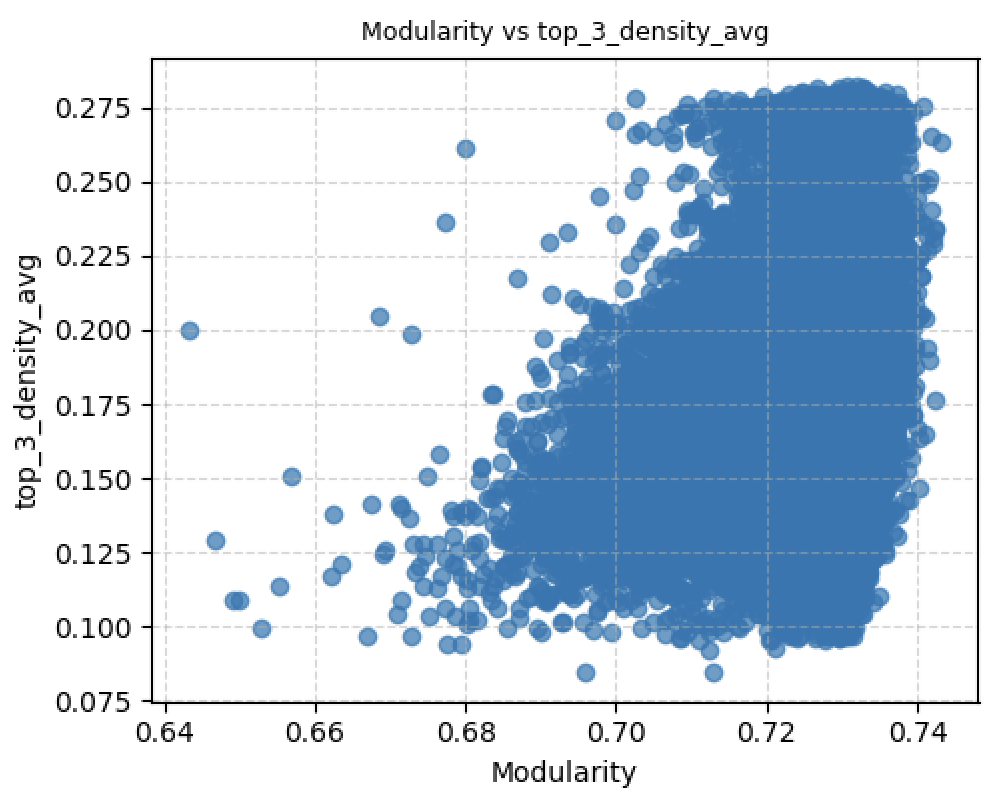}
        \caption{Modularity vs Top 3 Communities Density Avg}
        \label{fig:subplot_modopt2}
    \end{subfigure}
    \caption{Comparison of features correlations : YelpHotel Dataset}
    \label{fig:subplots_modularity_optimized}
\label{fig:corr}
\end{figure}

To study the correlation between the different properties, we computed scatter plot of all the solutions for each pair of metrics.  Figure~\ref{fig:corr} shows the solutions from the Genetic Algorithm which optimize the Modularity and the one that optimized Density for the YelpHotel dataset. A remarkable plot Figure~\ref{fig:subplot_modopt1} is show the number of communities and average cut size in modularity optimized solution. The plot mostly look hyperbolic. Upon inspection what happens is that most solution have about the same total cut size. But some solution have more tiny solution than other. So in this solution pool, the total cutsize does not change much, but the number of community being different, the denominator (the number of community) dominates the signal in the calculation. As such the plot look hyperbolic. This is why we introduced the average of the largest 3 communities. Other than that relation, the different properties have some correlation but it is obviously low. A lower top-3 conductance average correlate weakly correlate to a higher modularity. Similarly in Figure~\ref{fig:subplot_modopt2}, a higher modularity correlates to a higher top-3 density average.

The density optimized solution show similar patterns, the correlation is clearly stronger: Keeping community sizes smaller improves average density, decreases average cut size, and reduces conductance. The density optimized solution show stronger correlation than the modularity optimized one because the solution span a wider in range of values. In general, it seems that optimizing density with the genetic algorithm does not lead to good solutions in term of the other metrics.

We observe similar types of correlations in other function-optimized solutions for different application-based datasets as well (not shown). From the analysis, we can see that the properties are not quite correlated similarly or linearly and are rather complex. There is noise in the correlations that is difficult to understand at a glance. 

While there is some correlation in the solutions, even the solutions with the highest modularity still see a span in other metrics of 5\% to 10\%. So in the most optimized solution (which are the solution you would consider) the properties do not show a very strong correlation. In other words, optimizing one objective does not optimize the other ones.

\section{Community Structure Properties Impact on Task performance}
\label{sec:downstreamspace}

\begin{figure}[t]
    \centering    \includegraphics[width=0.8\linewidth]{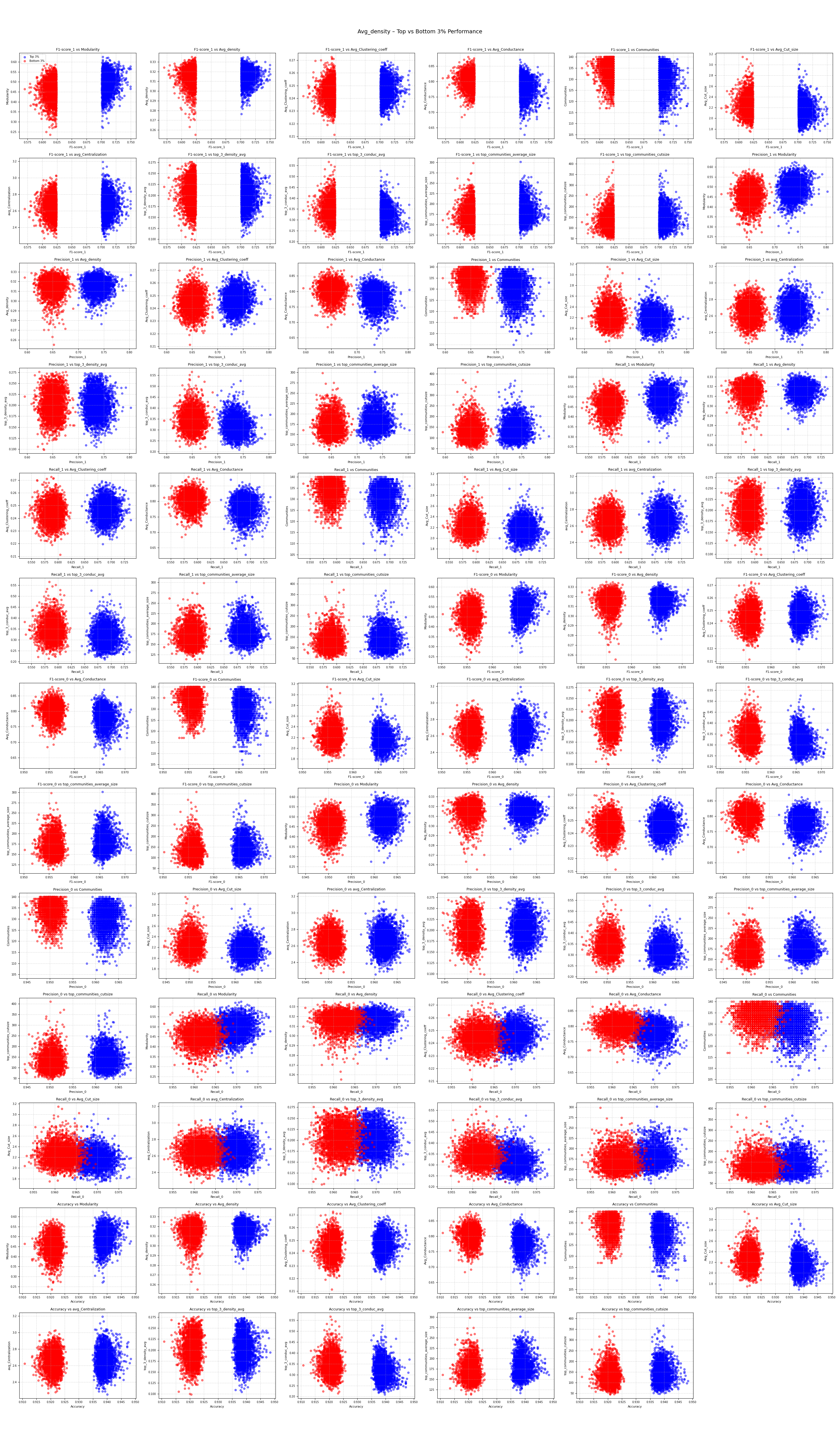}
    \caption{Anomaly Detection Prediction performance for Average Density optimized solutions: comparison of top and bottom 3\% cases: Yelp dataset}
    
    \label{fig:perf_yelp1}
\end{figure}
\begin{figure}[t]
    \centering    \includegraphics[width=0.8\linewidth]{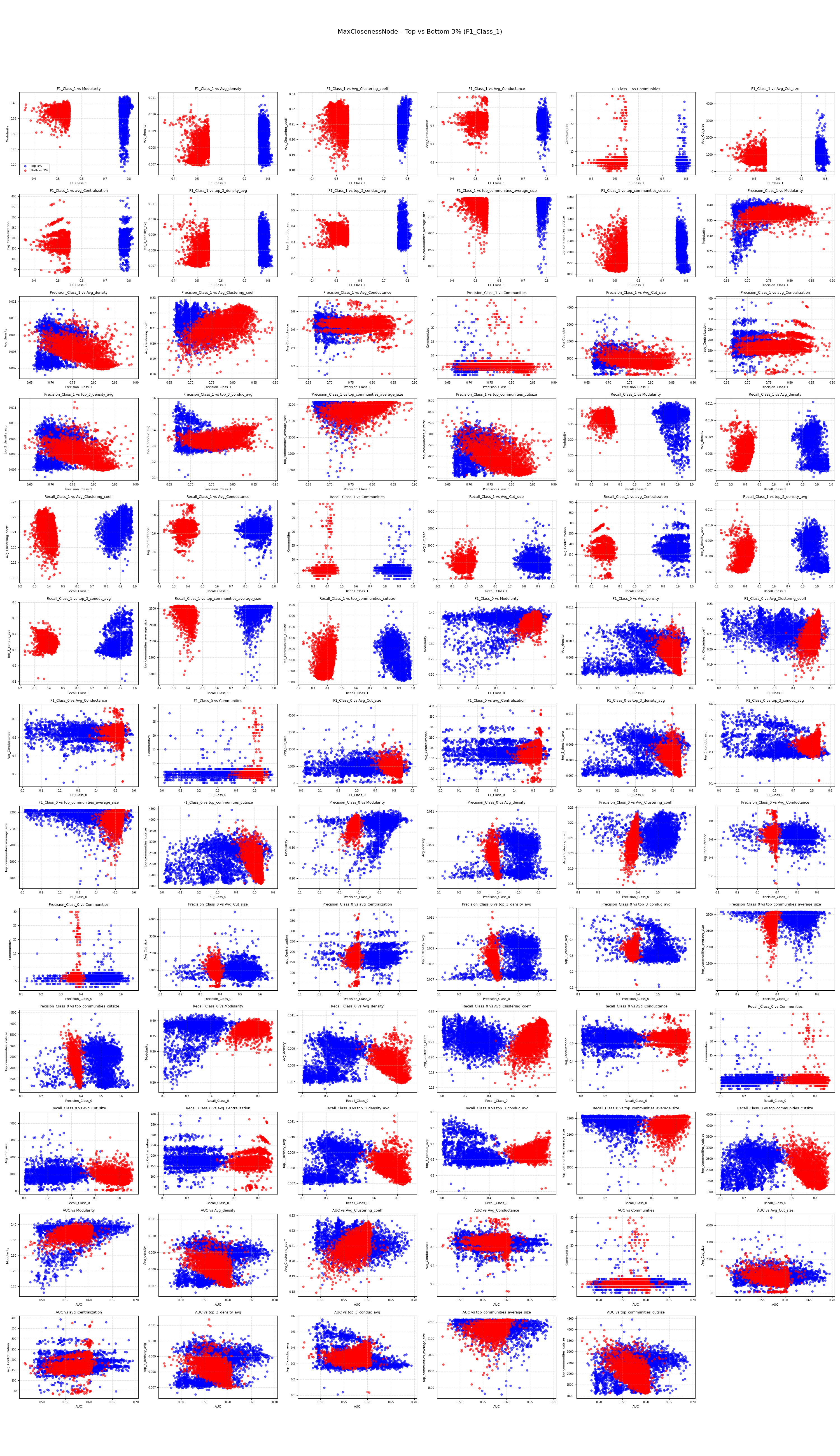}
    \caption{Trust prediction Performance (MaxClosenessNode Centrality) for Modularity Optimized Solutions: comparison of top and bottom 3\% cases.}
    
    \label{fig:perf_ciao}
\end{figure}

Since optimizing one objective does not optimize the other one, a natural question is to determine which objective should be optimized when performing community detection.
We reuse the community structures obtained with the Genetic Algorithm and we perform the prediction analysis on downstream tasks for each of those structures. For each application, we follow the same steps discussed in Sections~\ref{sec:app_anomaly} and~\ref{sec:app_trust} to apply downstream tasks to these solutions and generate predictions. 

The purpose of doing the prediction analysis is to find which community property correlates with task performance. To do this, we sort the prediction values in order against the F1-score of Class 1 and plot the top and bottom 3\% of prediction values against the community properties. For anomaly detection, we focus on the F1-score for anomaly class nodes, and for Trust Prediction, we focus on the F1-score for Trust relation class 1. We compute the scatter plots for each community property against each downstream performance metric. We plot the top and bottom prediction simultaneously in different colors to see whether specific community properties impact the prediction score. Figure~\ref{fig:perf_yelp1} shows the plots for anomaly detection, and Figure~\ref{fig:perf_ciao} shows the plot for trust prediction.

\begin{figure}[t]
    \centering
    \begin{subfigure}{0.48\linewidth}
        \centering
        \includegraphics[width=\linewidth]{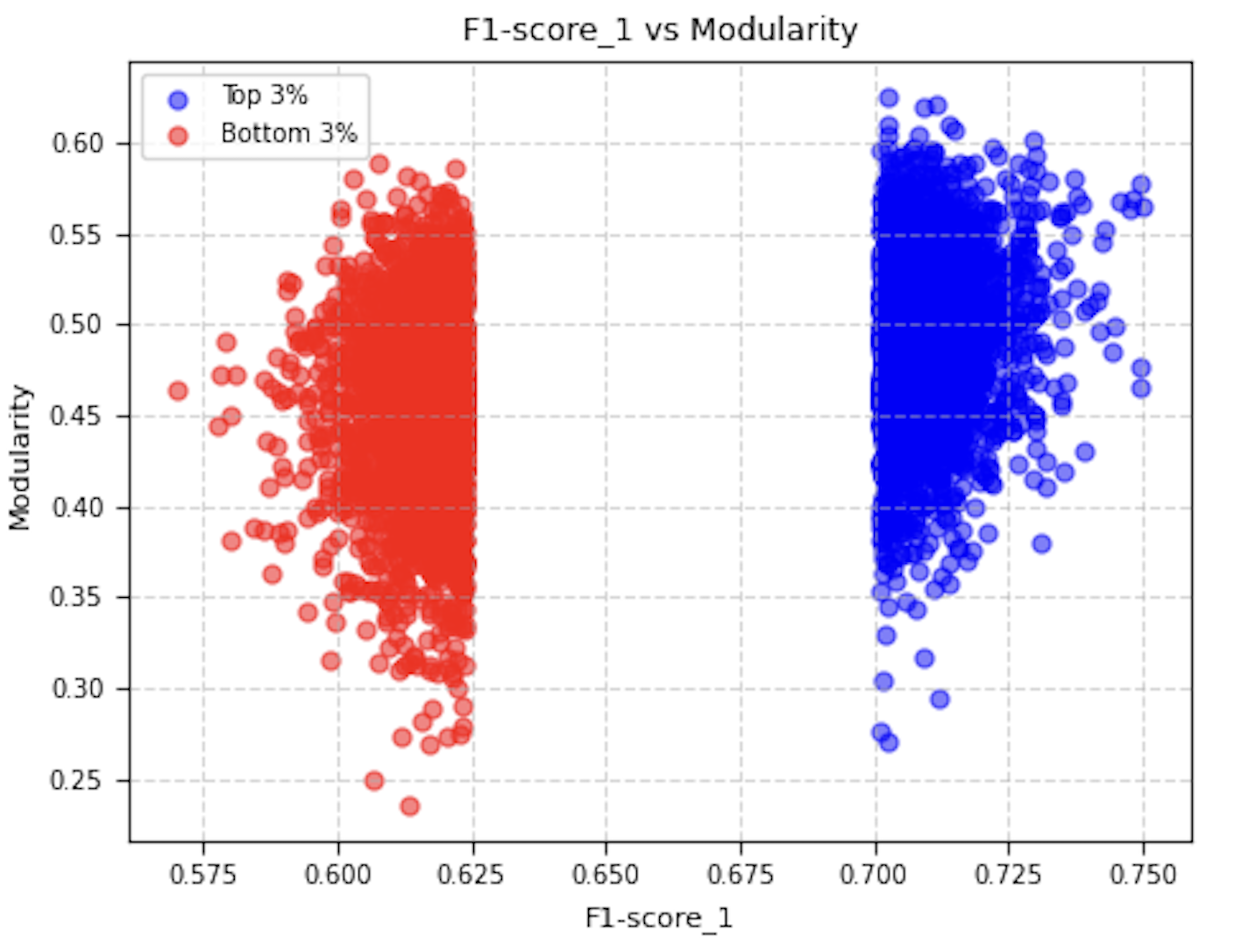}
        \caption{Modularity vs F1-score (Anomaly Class)}
        \label{fig:Modularity vs F1-score (Anomaly Class): Modularity Optimized }
    \end{subfigure}
    \hfill
    \begin{subfigure}{0.48\linewidth}
        \centering
        \includegraphics[width=\linewidth]{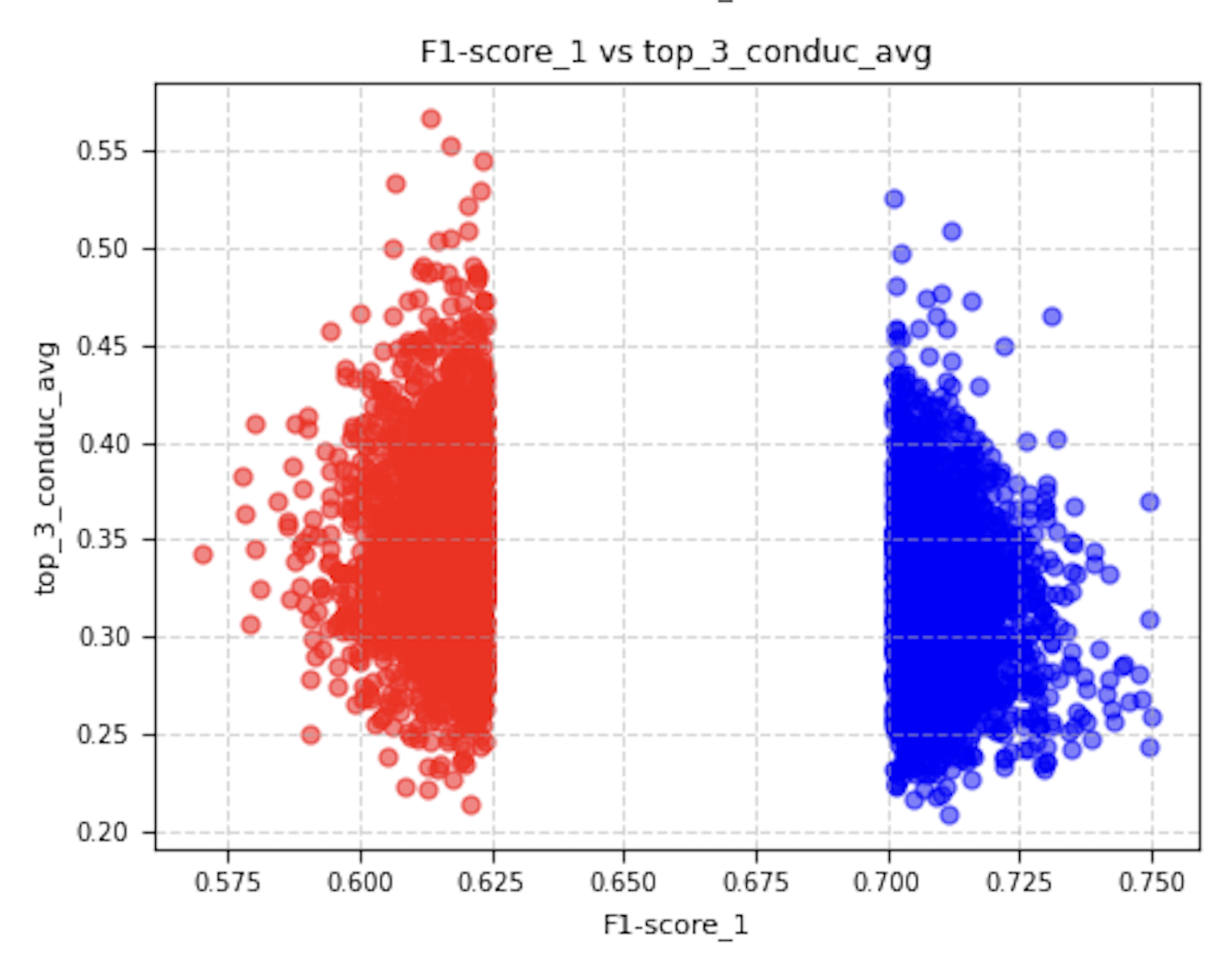}
        \caption{Top communities average conductance vs F1-score (Anomaly Class)}
        \label{fig:img2}
    \end{subfigure}
    \caption{Correlation plot of features and performance on Yelp dataset (Anomaly Detection)}
    \label{fig:feature_subplots_vs_perf_anomaly}
\end{figure}

Figure~\ref{fig:perf_yelp1} shows the results for anomaly detection. it reveals that certain structural properties of community-based solutions significantly influence task performance. In particular, in Figure~\ref{fig:feature_subplots_vs_perf_anomaly} shows higher modularity and lower conductance show higher F1-score, precision, and recall for the anomaly class (Class 1). Top-performing solutions tend to have higher modularity and lower conductance, indicating that well-separated and densely connected communities contribute positively to anomaly detection. These patterns are most prominent in the precision and F1-scores for Class 1. We observe similar correlation patterns between density and the clustering coefficient optimized solutions’ performance. 

\begin{figure}[t]
    \centering
    \begin{subfigure}{0.48\linewidth}
        \centering
        \includegraphics[width=\linewidth]{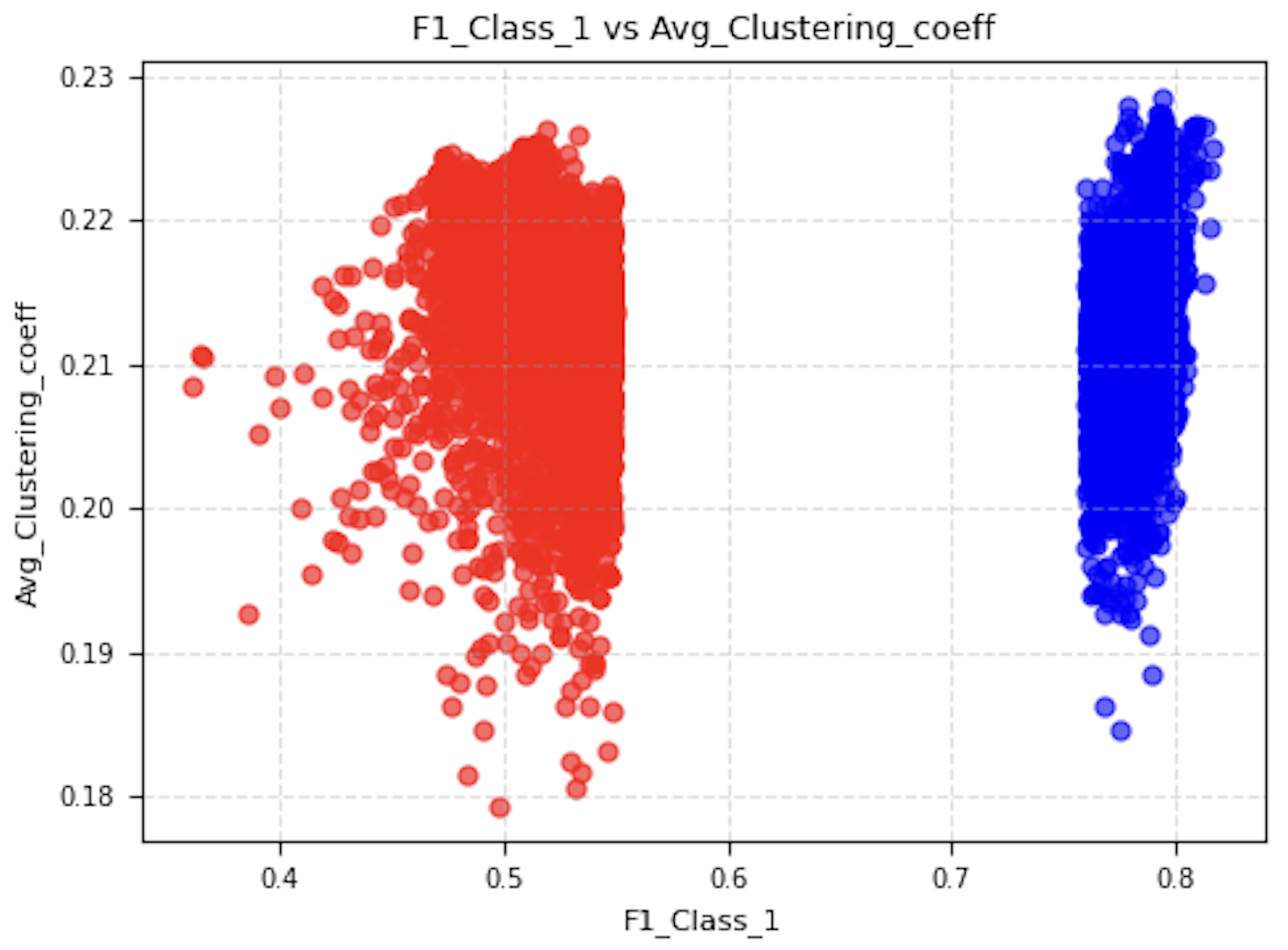}
        \caption{Avg clustering coefficient vs F1-score (Trust relation class)}
    \end{subfigure}
    \hfill
    \begin{subfigure}{0.48\linewidth}
        \centering
        \includegraphics[width=\linewidth]{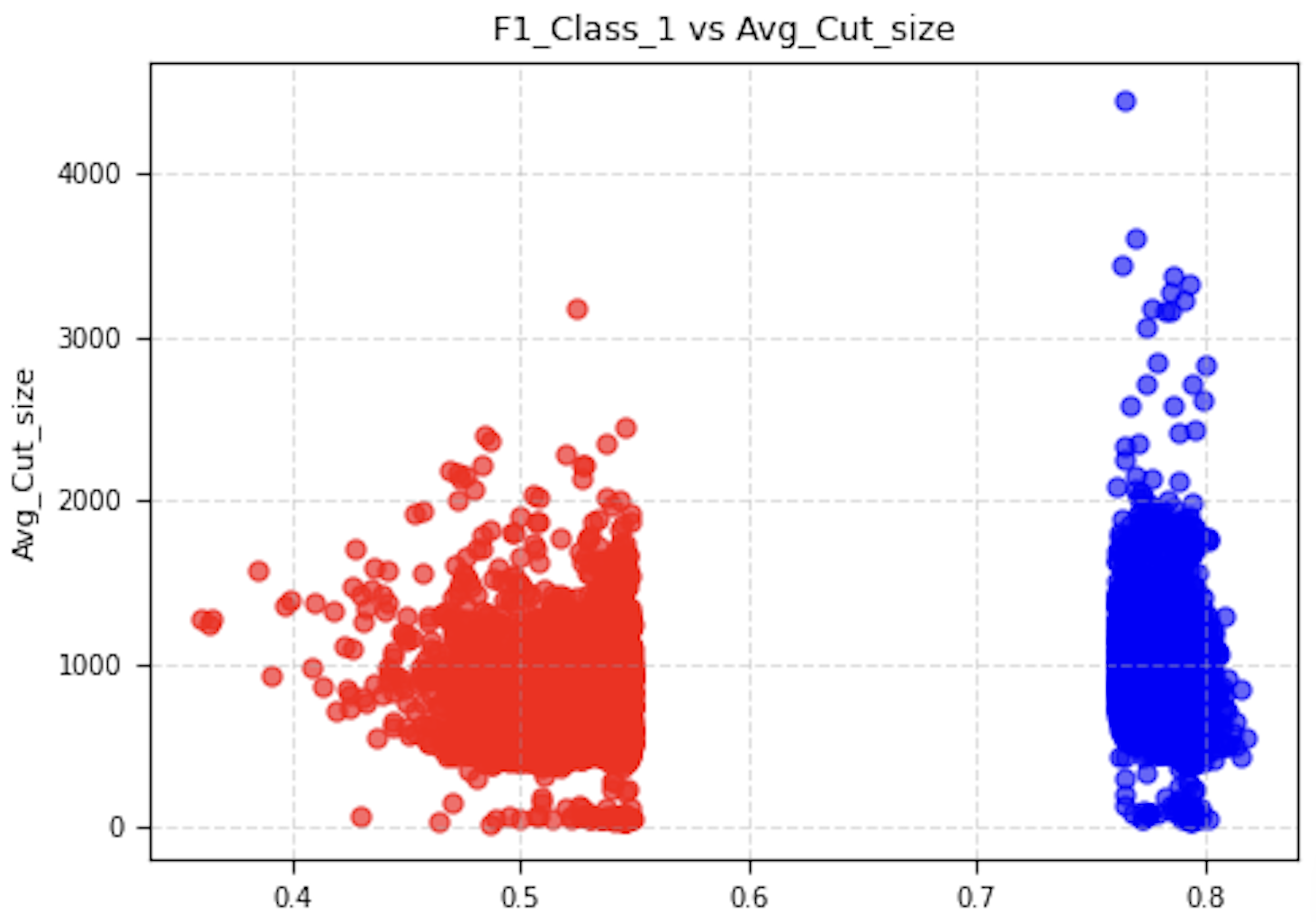}
        \caption{Avg cut size vs F1-score (Trust relation class)}
    \end{subfigure}
    \caption{Correlation plot of features and performance on Ciao dataset (Trust relation class)}
    \label{fig:feature_subplots_vs_perf_trust}
\end{figure}

Figure~\ref{fig:perf_ciao} shows performance is driven primarily by community cohesion and separation: particularly, Figure~\ref{fig:feature_subplots_vs_perf_trust}  shows higher average clustering coefficient and larger cut size both correlate with better Class-1 F1 and recall in Trust Prediction. Average density offers only a small boost (not affecting precision or AUC), while modularity and conductance show little to no relationship with any metric—including AUC—so they’re poor guides for trust prediction here. A moderate partitioning of about 5–7 communities tends to work best, suggesting that neither very coarse nor very fine splits are optimal.

While there is a difference in the distribution of task performance score between (for instance) higher modularity and lower modularity solutions, the distributions mostly overlap. In other words, while a higher modularity might statistically give you better performance, the distributions are wide enough that looking at a single high modularity community detection solution might not give you the best performance. This result motivates the main conclusion of this paper: A single community detection algorithm will not give a solution that will happen to optimize your downstream task.
\section{ Downstream Tasks Prediction using Feature-based Model}
\label{sec:perflearning}

While a single solution that optimize a single objective might not necessarily give the best downstream performance, there clear (albeit weak) correlation between community properties and the performance of downstream tasks. We hypothesize that this relation can be learned with machine learning models. We are going to train a predictor that will take community properties as features and that will predict the performance on the downstream task. 
We used LightGBM~\cite{ke2017lightgbm}, a gradient boosting machine learning model leveraging decision trees. 

Training LightGBM on all the community structures we have generated using the Genetic Algorithm would not be particularly realistic. One does not want to have to train ten thousands downstream tasks to pick the best community structure. Rather one would want to train using a small number of community structure and be able to extract the better community structures. We did not sample the community structures uniformly. Rather, we employed Reinforcement Learning for sampling the training set to evaluate predictions, and trained a new model solely on the smallest sample chosen by Reinforcement Learning model.  

We trained LightGBM based on both applications, and using the different pools of solution generated by the Genetic Algorithm when optimizing the different function features. From Table~\ref{tab:rl_model_performance_ad} we can see that with a small sample size which is less than 1\% of number of samples, we got a Root Mean Square Error value of 0.02. The span of performance values in the top 3\% solution on that application is 0.02, so in other word at that prediction accuracy, the model can differentiate between solutions in the top 3\% of solution and the rest of the solution pool. In other words, provided a large number of community structures, we can identify the solution with the highest prediction performance by training only a small fraction of downstream models. We tested the accuracy of the model when they are trained on a set of solution and tested on a different set of solution. Table~\ref{tab:rl_ad_perf_within} shows within the anomaly detection task, the RMSE of the model range between 0.02 and 0.07. Table~\ref{tab:rl_tp_perf_within} shows the trust prediction application sees worse performance ranging from 0.04 to 0.21. Though the bad performance is mostly due to clustering coefficient optimized solution which are deriving particularly bad downstream performance.

\begin{table}[t]
\centering

\caption{Prediction accuracy of LightGBM for the anomaly detection task}
\label{tab:rl_model_performance_ad}
\begin{tabular}{|l|c|c|c|}
\hline
\textbf{Model} & \textbf{Number of Samples} & \textbf{Fraction} & \textbf{RMSE} \\ \hline
Density-based Model            & 113 & 0.0012 & 0.0207 \\ \hline
Modularity-based Model         & 107 & 0.0012 & 0.0244\\ \hline
Clustering Coefficient-based Model & 107 & 0.0012 & 0.0197 \\ \hline
Conductance-based Model        & 144 & 0.0060 & 0.0190 \\ \hline
\end{tabular}
\end{table}

\begin{table}[htbp]
\centering
\caption{Anomaly Detection within domain RMSE (Trained vs Tested)}
\label{tab:rl_ad_perf_within}
\begin{tabular}{|l|c|c|c|c|}
\hline
\textbf{Trained on / Tested on} & \textbf{Density} & \textbf{Modularity} & \textbf{Clustering Coefficient} & \textbf{Conductance} \\
\hline
Density & 0.02066 & 0.04890 & 0.02047 & 0.02963 \\
\hline
Modularity & 0.051535 & 0.024604 & 0.056615 & 0.070284 \\
\hline
Clustering Coefficient & 0.02169 & 0.05402 & 0.01975 & 0.02566 \\
\hline
Conductance & 0.02921 & 0.06806 & 0.02303 & 0.01871 \\
\hline
\end{tabular}
\end{table}

\begin{table}[htbp]
\centering
\caption{Trust Prediction: within domain RMSE (Trained vs Tested)}
\label{tab:rl_tp_perf_within}
\begin{tabular}{|l|c|c|c|c|}
\hline
\textbf{Trained on / Tested on} & \textbf{Density} & \textbf{Modularity} & \textbf{Clustering Coefficient} & \textbf{Conductance} \\
\hline
Density      & 0.049720 & 0.075779 & 0.143525 & 0.052269 \\ \hline
Modularity   & 0.060626 & 0.040261 & 0.192591 & 0.068009 \\ \hline
Clustering Coeff. & 0.200288 & 0.236842 & 0.168242 & 0.210932 \\ \hline
Conductance  & 0.120378 & 0.130569 & 0.123680 & 0.120554 \\
\hline
\end{tabular}
\end{table}






\begin{table}[htbp]
\centering
\caption{Train on Anomaly detection and tested on Trust Prediction RMSE (Trained vs.\ Tested)}
\label{tab:cross_adtotp_rmse}
\begin{tabular}{|l|c|c|c|c|}
\hline
\textbf{Trained on / Tested on} & \textbf{Density} & \textbf{Modularity} & \textbf{Clustering Coefficient} & \textbf{Conductance} \\ \hline
Density                & 0.053372 & 0.067670 & 0.066661 & 0.051788 \\ \hline
Modularity             & 0.058379 & 0.039379 & 0.196771 & 0.077442 \\ \hline
Clustering Coefficient & 0.213657 & 0.240244 & 0.168183 & 0.205337 \\ \hline
Conductance            & 0.122539 & 0.136586 & 0.119747 & 0.119835 \\ \hline
\end{tabular}
\end{table}

Moreover, we test the prediction of the model  across applications. Table~\ref{tab:cross_adtotp_rmse} shows when we trained the predictor on anomaly detection sample and predicted the performance of the solution for trust prediction, we can see an RMSE error ranges between 0.04 to 0.23 and trust prediction to anomaly detection, we get an error between 0.01 to 0.09. This tells us that the properties that make a community structure good for an application might give insight on the properties that will perform well on an other application.

\section{Conclusion}

We investigate how the choice of community-detection algorithm influences downstream tasks. Because different algorithms optimize distinct objectives, it is difficult to pinpoint which specific structural property drives performance. Our results suggest that no single property (e.g., modularity, density, conductance) directly explains the outcomes; rather, performance emerges from interactions among multiple properties, possibly including unobserved factors. As such no single algorithm will likely perform best on downstream task. We believe that we will need to same the community structure space to find the best solution. This motivates a model-based approach: instead generating a high number of community structures and performing maybe downstream task evaluation, we can learn what makes a good community structure for a particular problem. We shows experimentally that we can do by sampling only few community decompositions. In future work, we will build and evaluate such meta-models—trained on features like number of communities, clustering coefficient, and cut size—to directly predict task performance and recommend methods or configurations without relying on any single algorithm.

\bibliographystyle{plain}
\bibliography{references}

%
\end{document}